\documentstyle[12pt,epsfig]{article}
\def\lbldef#1#2{\expandafter\gdef\csname #1\endcsname {#2}}

\def\href#1#2{#2}  
\begin{document}
\baselineskip=15.5pt
\pagestyle{plain}
\setcounter{page}{1}
%\renewcommand{\thefootnote}{\fnsymbol{footnote}}
%--------+---------+---------+---------+---------+---------+---------+
%Title page

\begin{titlepage}

\begin{flushright}
CERN-TH/2000-271\\
hep-th/0009082
\end{flushright}
\vspace{10 mm}

\begin{center}
{\Large BPS Solitons in M5-Brane Worldvolume Theory\\ 
with Constant Three-Form Field}

\vspace{5mm}

\end{center}

\vspace{5 mm}

\begin{center}
{\large Donam Youm\footnote{E-mail: Donam.Youm@cern.ch}}

\vspace{3mm}

Theory Division, CERN, CH-1211, Geneva 23, Switzerland

\end{center}

\vspace{1cm}

\begin{center}
{\large Abstract}
\end{center}

\noindent

We study BPS solutions for a self-dual string and a neutral string in 
M5-brane worldvolume theory with constant three-form field.  We further  
generalize such solitons to superpose with a calibrated surface.  We 
also study a traveling wave on a calibrated surface in the constant 
three-form field background.

\vspace{1cm}
\begin{flushleft}
CERN-TH/2000-271\\
September, 2000
\end{flushleft}
\end{titlepage}
\newpage

\section{Introduction}

Recently, noncommutative theories have received renewed interest, after 
it is founded out that noncommutative theories can be naturally realized 
in M-theory compactified in the presence of constant three-form field 
background \cite{cds} and within string theories as the worldvolume theory 
of D-brane with nonzero constant NS $B$-field \cite{dh,sw}.  In the case 
of D-brane with constant magnetic NS $B$-field, one can take a decoupling 
limit of the D-brane worldvolume theory to achieve a noncommutative field 
theory with space/space noncommutativity.  However, in the case of electric 
NS $B$-field background, one cannot take a zero slope limit in such a way 
that the noncommutativity parameter is nonzero to obtain a field theory 
with space/time noncommutativity, since the electric NS $B$ field cannot 
be scaled to infinity due to its critical value beyond which the open 
string parameters are not well-defined \cite{sst}.  It is conjectured 
\cite{sst,br} that nearby the critical electric $B$ field with an appropriate 
scaling limit a novel noncritical open string theory, called noncommutative 
open string (NCOS) theory, which decouples from closed strings (and 
therefore also gravity) in the bulk, emerges.  It is conjectured 
\cite{om,bbs2} that the strong coupling limit of the NCOS theory in 
$4+1$ dimensions is the so-called OM theory, which is the decoupled (from 
gravity) theory of light open M2-brane ending on M5-brane in the constant 
three-form field background.  

It would be therefore interesting to study worldvolume solitons of the 
M5-brane with constant 3-form field.  The solitons of the brane worldvolume 
theories are interpreted as the intersections of the interesting branes 
\cite{pt}, with the intersection being the source for the charge 
carried by the worldvolume soliton \cite{str,tow,bs}.  For example, the 
M5-brane worldvolume soliton counterpart to the open M2-brane ending on 
M5-brane is the self-dual string \cite{hlw} and the three-brane soliton 
\cite{hlw2} on the M5-brane worldvolume is interpreted as the three-brane 
intersection of intersecting two M5-branes. The solitons in the D-brane 
worldvolume theory, i.e. BI solitons or BIons \cite{bi1,bi2,hlw}, with 
constant NS $B$ field were previously studied, for example, in Refs. 
\cite{ha1,ha2,ha3,mat,ha4,mor}.  In the case of the zero-brane soliton, 
corresponding to fundamental or D string ending on a D-brane, force on the 
endpoint due to nonzero constant NS $B$ field is shown \cite{ha1} to cause 
the string to be tilted.  The non-locality of the suspended string due to 
such tilting is in accordance with the fact that the D-brane worldvolume is 
noncommutative, which leads to uncertainty in measurement.  

In this paper, we study the BPS self-dual string and neural (or instanton) 
string solitons in the M5-brane worldvolume theory with constant three-form 
field, and their generalization on calibrated surfaces.  We also show that 
traveling wave on the calibrated surface with constant three-form field 
preserves supersymmetry.  In studying such worldvolume solitons, we follow 
the covariant equations of motion approach \cite{hs1,hs2,hs3} of M5-brane 
theory.

\section{Aspects of M5-Brane Worldvolume Theory}

We discuss the relevant aspect of the M5-brane worldvolume theory 
\cite{hs1,hs2,hs3,cal1,cal2,glw} for the purpose of fixing and 
defining notation.  The convention for indices is as follows.  The indices 
for the target space are underlined.  The indices from the beginning [middle] 
of the alphabet refer to the tangent space [coordinate space] indices.  The 
Latin [Greek] indices are for the bosonic [fermionic] coordinates.  The 
primed indices are for directions normal to the M5-brane worldvolume.  
So, the coordinates of the target and the worldvolume superspaces are 
respectively $Z^{\underline{M}}=(X^{\underline{m}},\Theta^{\underline{\mu}})$ 
and $z^M=(x^m,\theta^{\mu})$, and, for example, the target space index 
$\underline{a}$ [$\underline{\alpha}$] is decomposed as $\underline{a}=
(a,a^{\prime})$ with $a=0,1,...,5$ and $a^{\prime}=1^{\prime},...,5^{\prime}$ 
[$\underline{\alpha}=(\alpha,\alpha^{\prime})$ with $\alpha=1,...,16$ and 
$\alpha^{\prime}=1^{\prime},...,16^{\prime}$].  The fermionic indices $\alpha$ 
and $\alpha^{\prime}$, running from 1 to 16, are alternatively written 
respectively as $\alpha\,i$ and $^{\alpha}_i$ when appearing as subscript, 
and respectively as $\alpha\,i$ and $^i_{\alpha}$ when appearing as 
superscript, with $\alpha$ and $i$ running from 1 to 4.  For example, 
$\Theta^{\alpha^{\prime}}\to\Theta^i_{\alpha}$, $\Theta^{\alpha}\to
\Theta^{\alpha i}$ and $u^{\ \beta^{\prime}}_{\alpha}\to u^{\ \ \,j}_{\alpha 
i\beta}$.  We denote the vielbeins of the target and the M5-brane 
worldvolume superspaces as $E^{\ \underline{A}}_{\underline{M}}$ and 
$E^{\ A}_M$, respectively.  We denote the embedding matrix 
$\tilde{E}^{\ \underline{A}}_A=E^{\ M}_A\partial_MZ^{\underline{M}}E^{\ 
\underline{A}}_{\underline{M}}$ with a tilde to avoid confusion with 
vielbeins.  
The worldvolume theory of the M5-brane is described by the $(2,0)$ tensor 
multiplet containing five scalars $X^{a^{\prime}}$, sixteen fermions 
$\Theta^{\alpha^{\prime}}$ and a self-dual field strength $h_{abc}$.  Here, 
$X^{m^{\prime}}$ and $\Theta^{\alpha^{\prime}}$ are identified as the 
transverse components of the target superspace coordinates $(X^{\underline{m}},
\Theta^{\underline{\mu}})$.  

In this paper, we consider an M5-brane embedded in flat eleven-dimensional 
target space with the metric $\hat{g}_{\underline{m}\underline{n}}=
\eta_{\underline{m}\underline{n}}$.  Just as in the case of D-brane with 
the NS $B$-field, the induced metric $g_{mn}=\eta_{\underline{m}\underline{n}}
\partial_mX^{\underline{m}}\partial_nX^{\underline{n}}$ on the M5-brane 
worldvolume does not correspond to the metric on the M5-brane felt by the 
open M2-brane in the presence of the background three-form field strength 
${\cal H}_{mnp}$.  Here, the gauge invariant field strength ${\cal H}_{mnp}$ 
of the M5-brane worldvolume two-form potential $b_{mn}$ given by
\begin{equation}
{\cal H}_{mnp}=\partial_{[m}b_{np]}+\partial_{m}X^{\underline{m}}
\partial_nX^{\underline{n}}\partial_pX^{\underline{p}}
C_{\underline{m}\underline{n}\underline{p}},
\label{ginvfs}
\end{equation}
where $C_{\underline{m}\underline{n}\underline{p}}$ is the three-form 
potential in the eleven-dimensional supergravity, is related to the self-dual 
field strength $h_{abc}$ as 
\begin{equation}
{\cal H}_{mnp}=E^{\ a}_mE^{\ b}_nE^{\ c}_pm^{\ d}_bm^{\ e}_c
h_{ade}=e^{\ a}_me^{\ b}_ne^{\ c}_p(m^{-1})^{\ e}_ch_{abe}. 
\label{relbetwnhH}
\end{equation}
The self-duality condition on $h_{abc}$, $h_{abc}={1\over 6}\epsilon_{abcdef}
h^{def}$, is translated to the following non-linear self-duality condition on 
${\cal H}_{mnp}$:
\begin{equation}
{\sqrt{-{\rm det}\,g}\over 6}\epsilon_{mnpqrs}{\cal H}^{qrs}=
{{1+K}\over 2}(G^{-1})^{\ r}_m{\cal H}_{npr},
\label{selfdual}
\end{equation}
where $K\equiv\sqrt{1+{1\over{24}}{\cal H}_{mnp}g^{mq}g^{nr}g^{ps}
{\cal H}_{qrs}}$.  
It is suggested in Ref. \cite{bbs1} that it is the metric $G_{mn}=E^{\ a}_m
E^{\ b}_n\eta_{ab}$ associated with $E^{\ a}_m$ that is felt by an open 
M2-brane ending on the M5-brane.  The vielbein $e^{\ a}_m$ associated with 
$g_{mn}$ is related to $E^{\ a}_m$ as $E^{\ a}_m=e^{\ b}_m
(m^{-1})^{\ a}_b$, where $m^{\ b}_a\equiv\delta^{\ b}_a-2h_{acd}h^{bcd}$.  
So, the open M2-brane metric $G_{mn}$ is expressed in terms of $g_{mn}$ and 
${\cal H}_{mnp}$ as
\begin{equation}
G_{mn}={{1+K}\over{2K}}\left(g_{mn}+{1\over 4}{\cal H}_{mpq}
g^{pr}g^{qs}{\cal H}_{nrs}\right).
\label{openmemmet}
\end{equation}
As expected, the open M2-brane metric $G_{mn}$ becomes $g_{mn}$ when 
${\cal H}_{mnp}=0$ and reduces to the open string metric \cite{sw} on a 
D-brane upon double dimensional reduction.  

The equations of motion of the M5-brane can be obtained by analyzing the 
torsion equation subject to the superembedding condition $\tilde{E}^{\ 
\underline{a}}_{\alpha}=0$.  In a flat target superspace background in the 
static gauge, where the fermionic field $\Theta^{\underline{\alpha}}=
(\Theta^{\alpha},\Theta^{\alpha^{\prime}})$ is zero and $X^m=x^m$, the 
bosonic equations of motion are
\begin{eqnarray}
G^{mn}\nabla_m\nabla_nX^{a^{\prime}}&=&0,
\cr
G^{mn}\nabla_m{\cal H}_{npq}&=&0,
\label{boseqs}
\end{eqnarray}
where the connection in the covariant derivative $\nabla_m$ is defined 
from $g_{mn}$, i.e. $\Gamma^{\ \ \ p}_{mn}=\partial_m\partial_n
X^{a^{\prime}}\partial_qX^{b^{\prime}}g^{pq}\delta_{a^{\prime}b^{\prime}}$.  

We are particularly interested in solutions that preserve part of 
supersymmetry.  The condition that the static gauge condition 
$\Theta^{\alpha}=0$ be preserved under the combined rigid supersymmetry 
transformation of a flat target superspace and super-reparametrization of 
the worldvolume induced on the target superspace leads to the following 
form of the supersymmetry transformation for the fermions:
\begin{equation}
\delta\Theta^{\alpha^{\prime}}=-\epsilon^{\alpha}
(\tilde{E}^{-1})^{\ \beta}_{\alpha}\tilde{E}^{\ \alpha^{\prime}}_{\beta},
\label{susytran}
\end{equation}
where the non-linearly realized symmetry parameterized by 
$\epsilon^{\alpha^{\prime}}$ is set to zero.  Making use of the 
projection operators $(\tilde{E}^{-1})^{\ \,\beta}_{\underline{\alpha}}
\tilde{E}^{\ \,\underline{\gamma}}_{\beta}={1\over 2}(1+\Gamma)^{\ \,
\underline{\gamma}}_{\underline{\alpha}}$ and $(\tilde{E}^{-1})^{\ \,
\beta^{\prime}}_{\underline{\alpha}}\tilde{E}^{\ \,
\underline{\gamma}}_{\beta^{\prime}}={1\over 2}(1-\Gamma)^{\ \,
\underline{\gamma}}_{\underline{\alpha}}$, one can put the variation 
(\ref{susytran}) for $\Theta^{\alpha^{\prime}}$ into the form:
\begin{equation}
\hat{\delta}\Theta^{\alpha^{\prime}}\equiv\delta\Theta^{\gamma^{\prime}}
\left({{1-\Gamma}\over 2}\right)^{\ \,\alpha^{\prime}}_{\gamma^{\prime}}=
-{1\over 2}\epsilon^{\gamma}\Gamma^{\ \,\alpha^{\prime}}_{\gamma}.
\label{susytran2}
\end{equation}
This has to be set to zero to make the condition $\Theta^{\alpha^{\prime}}=0$ 
to be invariant under the supersymmetry transformation.  
In terms of the bosonic fields of the M5-brane worldvolume theory, the 
supersymmetry variation (\ref{susytran2}) is expressed as
\begin{eqnarray}
\hat{\delta}\Theta^{\ \,j}_{\beta}&=&-{1\over 2}\epsilon^{\alpha i}\left[
{\rm det}(e^{-1})\partial_mX^{a^{\prime}}(\gamma^m)_{\alpha\beta}
(\gamma_{a^{\prime}})^{\ \,j}_i\right.
\cr
& &-{1\over{3!}}{\rm det}(e^{-1})\partial_{m_1}X^{a^{\prime}_1}
\partial_{m_2}X^{a^{\prime}_2}\partial_{m_3}X^{a^{\prime}_3}
(\gamma^{m_1m_2m_3})_{\alpha\beta}(\gamma_{a^{\prime}_1a^{\prime}_2
a^{\prime}_3})^{\ \,j}_i
\cr
& &+{1\over{5!}}{\rm det}(e^{-1})\partial_{m_1}X^{a^{\prime}_1}
\cdots\partial_{m_5}X^{a^{\prime}_5}(\gamma^{m_1\dots m_5})_{\alpha\beta}
(\gamma_{a^{\prime}_1\dots a^{\prime}_5})^{\ \,j}_i
\cr
& &-h^{m_1m_2m_3}\partial_{m_2}X^{a^{\prime}_2}\partial_{m_3}
X^{a^{\prime}_3}(\gamma_{m_1})_{\alpha\beta}(\gamma_{a^{\prime}_2
a^{\prime}_3})^{\ \,j}_i
\cr
& &\left.-{1\over 3}h^{m_1m_2m_3}(\gamma_{m_1m_2m_3})_{\alpha
\beta}\delta^{\ \,j}_i\right],
\label{susyvar}
\end{eqnarray}
where $\gamma_m=\delta^a_m\gamma_a$.  

It is the purpose of this paper to study solitons in the M5-brane worldvolume 
theory in the constant three-form field ${\cal H}$ background.  When 
${\cal H}$ is constant, up to a Lorentz transformation the nonzero components 
of ${\cal H}$ satisfying the non-linear self-duality condition 
(\ref{selfdual}) are ${\cal H}_{012}$ and ${\cal H}_{345}$ \cite{sw,bbs1}.  
Equivalently, nonzero components of the worldvolume field $h_{abc}$ are
\begin{equation} 
h_{012}=-h_{345}=h={\rm constant}, 
\label{bdcdtn}
\end{equation}
which corresponds to ${\cal H}_{012}={h\over{1+4h^2}}\equiv{1\over 4}
\sin\theta$ and ${\cal H}_{345}=-{h\over{1-4h^2}}=-{1\over 4}\tan\theta$, 
according to Eq. (\ref{relbetwnhH}).  In the infinite momentum frame 
(boosted along the $x^5$-direction), the nonzero components of $h_{abc}$ are
\begin{equation}
h_{012}=-h_{034}=-h_{512}=h_{534}={\rm constant}.
\label{newbdcdtn}
\end{equation} 
The derivation of M5-brane solitons in the constant ${\cal H}$ background 
is along the same line as the case with zero ${\cal H}$ background 
\cite{hlw,hlw2,glw}, except that one has to impose the boundary condition 
(\ref{bdcdtn}) or (\ref{newbdcdtn}) at the infinity of the worldvolume.

\section{Self-Dual String}

The self-dual string in the M5-brane worldvolume theory is interpreted as the 
boundary of a M2-brane ending on a M5-brane:
\begin{equation}
\matrix{M5:&1&2&3&4&5& \cr M2:&1& & & & &6}
\label{m2m5}
\end{equation} 
All fields of the self-dual string soliton are independent of the worldvolume 
coordinates $(x^0,x^1)$ of the string soliton.  We denote the four M5-brane 
worldvolume indices for the directions transverse to the string soliton with 
tilde, i.e. $\tilde{a},\tilde{m}=2,...,5$.  We let only one of the scalar 
fields, which we choose $X^{1^{\prime}}\equiv\phi$, to be active.  The bosonic 
worldvolume field Ansatz for the string soliton is
\begin{equation}
X^{1^{\prime}}=\phi,\ \ \ \ \ \ \ 
h_{01\tilde{a}}=v_{\tilde{a}},\ \ \ \ \ \ \ 
h_{\tilde{a}\tilde{b}\tilde{c}}=\epsilon_{\tilde{a}\tilde{b}\tilde{c}\tilde{d}}
v^{\tilde{d}},
\label{bosantz1}
\end{equation}
with the remaining components of $h_{abc}$ vanishing, along with the 
boundary condition $h_{01\tilde{a}}=h\delta^2_{\tilde{a}}$ at infinity.  

Substituting the field Ansatz (\ref{bosantz1}) into Eq. (\ref{susyvar}),  
one obtains the following supersymmetry variation of fermions $\Theta$:
\begin{equation}
\hat{\delta}\Theta=-{1\over 2}\epsilon\left[{\rm det}(e^{-1})\partial_m
\phi\gamma^m\gamma_{1^{\prime}}-2\gamma^{01}\left\{v^{\tilde{m}}
\gamma_{\tilde{m}}+{\rm det}(e^{-1})v_{\tilde{m}}\gamma^{\tilde{m}}
\right\}\right].
\label{susyvarselfstr}
\end{equation}
If one requires spinors to satisfy the following constraint
\begin{equation}
\epsilon=\epsilon\gamma^{01}\gamma_{1^{\prime}},
\label{sustconstr}
\end{equation}
then from the Killing spinor equation $\hat{\delta}\Theta=0$ one obtains 
the following Bogomol'nyi condition on the fields:
\begin{equation}
v_{\tilde{a}}={1\over 2}{{\delta^{\tilde{m}}_{\tilde{a}}\partial_{\tilde{m}}
\phi}\over{1+\sqrt{1+(\partial_{\tilde{m}}\phi)^2}}},
\label{bogrelflds}
\end{equation}
where $(\partial_{\tilde{m}}\phi)^2\equiv\delta^{\tilde{m}\tilde{n}}
\partial_{\tilde{m}}\phi\partial_{\tilde{n}}\phi$.

Making use of Eq. (\ref{relbetwnhH}) along with the field Ansatz 
(\ref{bosantz1}), one obtains the following nonvanishing components of  
${\cal H}$:
\begin{eqnarray}
{\cal H}_{01\tilde{m}}&=&{1\over{1+4v^2}}e^{\ \tilde{a}}_{\tilde{m}}
v_{\tilde{a}},
\cr
{\cal H}_{\tilde{m}\tilde{n}\tilde{p}}&=&{\sqrt{1+(\partial_{\tilde{m}}\phi)^2}
\over{1-4v^2}}\epsilon_{\tilde{m}\tilde{n}\tilde{p}\tilde{q}}
e^{\ \tilde{q}}_{\tilde{a}}v^{\tilde{a}},
\label{nonvanh1}
\end{eqnarray}
where $v^2\equiv\delta^{\tilde{a}\tilde{b}}v_{\tilde{a}}v_{\tilde{b}}$.  
The vielbein $e^{\ \,a}_m$ associated with the induced metric $g_{mn}$ 
on the M5-brane worldvolume is given by $(e^{\ \,a}_m)={\rm diag}
(1,1,e^{\ \,\tilde{a}}_{\tilde{m}})$ with $e^{\ \tilde{a}}_{\tilde{m}}=
\delta^{\tilde{a}}_{\tilde{m}}+c\partial_{\tilde{m}}\partial^{\tilde{a}}
\phi$, where $c\equiv(-1+\sqrt{1+\delta^{\tilde{m}\tilde{n}}
\partial_{\tilde{m}}\phi\partial_{\tilde{n}}\phi})/\delta^{\tilde{m}\tilde{n}}
\partial_{\tilde{m}}\phi\partial_{\tilde{n}}\phi$.
These nonzero components of ${\cal H}$ simplify to the following forms 
after the Bogomol'nyi condition (\ref{bogrelflds}) is substituted:
\begin{eqnarray}
{\cal H}_{01\tilde{m}}&=&{1\over 4}\partial_{\tilde{m}}\phi,
\cr
{\cal H}_{\tilde{m}\tilde{n}\tilde{p}}&=&{1\over 4}
\epsilon_{\tilde{m}\tilde{n}\tilde{p}\tilde{q}}\delta^{\tilde{q}\tilde{r}}
\partial_{\tilde{r}}\phi,
\label{nonvanh2}
\end{eqnarray}
where $\epsilon_{\tilde{m}\tilde{n}\tilde{p}\tilde{q}}=
e^{\ \,\tilde{a}}_{\tilde{m}}e^{\ \,\tilde{b}}_{\tilde{n}}
e^{\ \,\tilde{c}}_{\tilde{p}}e^{\ \,\tilde{d}}_{\tilde{q}}
\epsilon_{\tilde{a}\tilde{b}\tilde{c}\tilde{d}}$.  

To find the expression for the scalar $\phi$, one has to solve the equation 
(\ref{boseqs}) for the scalar $X^{5^{\prime}}=\phi$.  It is shown 
\cite{blw} that generally the equation of motion $G^{mn}\nabla_m\nabla_n
X^{a^{\prime}}=0$ for the scalar $X^{a^{\prime}}$ implies $G^{mn}\partial_m
\partial_nX^{a^{\prime}}=0$.  Note, the worldvolume fields for self-dual 
string solution are independent of the worldvolume coordinates $x^0$ and 
$x^1$.  And it can be shown by applying Eq. (\ref{bogrelflds}) that 
$G^{\tilde{m}\tilde{n}}\propto\delta^{\tilde{m}\tilde{n}}$.  So, the scalar 
$\phi$ satisfies the flat Laplace's equation:
\begin{equation}
\delta^{\tilde{m}\tilde{n}}\partial_{\tilde{m}}\partial_{\tilde{n}}\phi=0.
\label{fllapphi}
\end{equation}
From Eq. (\ref{nonvanh2}), one can see that the solution to Eq. 
(\ref{fllapphi}), satisfying the boundary condition ${\cal H}_{012}=
{h\over{1+4h^2}}={1\over 4}\sin\theta$ at infinity, that describes array of 
strings with charge $Q_K$ located at $x^{\tilde{m}}=y^{\tilde{m}}_K$ is 
given by
\begin{equation}
\phi=\phi_0+\sum_K{{2Q_K}\over{|x-y_K|^2}}+x^2\sin\theta.
\label{phisol}
\end{equation}
Due to the self-duality condition on ${\cal H}$, the string at $x^{\tilde{m}}
=y^{\tilde{m}}_K$ carries the same electric and magnetic charges 
$Q_E=Q_M=Q_K$.  From the second equation of Eq. (\ref{nonvanh1}) or 
(\ref{nonvanh2}), one can see that the boundary condition ${\cal H}_{345}=
-{h\over{1-4h^2}}=-{1\over 4}\tan\theta$ at infinity is automatically 
satisfied by Eq. (\ref{phisol}).  From the expression (\ref{phisol}) for the 
scalar $\phi$, one can see that the M2-brane is tilted towards the 
$x^2$-direction due to the force felt by the self-dual string at the boundary 
of the suspended M2-brane in the background of constant ${\cal H}$ field.  
This force due to the constant ${\cal H}$ field is canceled by the tension 
of the M2-brane.  

By compactifying the above self-dual string solution along the 
$x^1$-direction, one obtains the following 0-brane soliton (BIon) on the 
D4-brane worldvolume in the constant $B$ field background studies in 
Refs. \cite{ha1,ha2}:
\begin{equation}
{\cal F}_{0\tilde{m}}={1\over 4}\partial_{\tilde{m}}\phi,\ \ \ \ \ 
\phi=\phi_0+\sum_K{{2Q_K}\over{|x-y_K|^2}}+x^2\sin\theta,
\label{bionbfld}
\end{equation}
where ${\cal F}_{\tilde{m}\tilde{n}}\equiv{\cal H}_{\tilde{m}\tilde{n}1}$.  
Dimensional reduction along, say, the $x^5$-direction leads to the 
following string soliton on the D4-brane worldvolume:
\begin{equation}
{\cal F}_{\tilde{m}\tilde{n}}={1\over 4}\epsilon_{\tilde{m}\tilde{n}\tilde{p}}
\delta^{\tilde{p}\tilde{q}}\partial_{\tilde{q}}\phi,\ \ \ \ \ 
\phi=\phi_0+\sum_K{{2Q_K}\over{|x-y_K|^2}}+x^2\sin\theta,
\label{stringd4sol}
\end{equation}
where ${\cal F}_{\tilde{m}\tilde{n}}={\cal H}_{\tilde{m}\tilde{n}5}$.  

It is straightforward to show that a self-dual string on a calibrated 
surface in the constant $h_{abc}$ field background also preserves 
supersymmetry.  The amount of supersymmetry preserved depends on the 
type of calibrated surface.  We denote $X^I$ as the scalars associated 
with the calibrated surface and, as above, $X^{1^{\prime}}=\phi$ is the 
scalar of the self-dual string soliton.  The supersymmetry variation of 
the fermion is given by
\begin{eqnarray}
\hat{\delta}\Theta&=&-{1\over 2}\epsilon\left[\left\{{\rm det}(e^{-1})
\partial_mX^I\gamma^m\gamma_I\right.\right.
\cr
& &\ \ \ -{1\over{3!}}{\rm det}(e^{-1})\partial_{m_1}X^{I_1}
\partial_{m_2}X^{I_2}\partial_{m_3}X^{I_3}\gamma^{m_1m_2m_3}
\gamma_{I_1I_2I_3}
\cr
& &\ \ \ \left.+{1\over{5!}}{\rm det}(e^{-1})\partial_{m_1}X^{I_1}\cdots
\partial_{m_5}X^{I_5}\gamma^{m_1...m_5}\gamma_{I_1...I_5}\right\}
\cr
& &\ \ \ \left.+{\rm det}(e^{-1})\partial_m\phi\gamma^m\gamma_{1^{\prime}}
-2\gamma^{01}\left(v^{\tilde{m}}\gamma_{\tilde{m}}+{\rm det}(e^{-1})
v_{\tilde{m}}\gamma^{\tilde{m}}\right)\right].
\label{calsfdsusyvar}
\end{eqnarray}
The self-dual string soliton on the calibrated surface preserves 
supersymmetry, if it satisfies the Killing spinor equation $\hat{\delta}
\Theta=0$ with nonzero $\epsilon$.  We rather consider two equations obtained 
by setting two separate terms, i.e. terms in the curly bracket of Eq. 
(\ref{calsfdsusyvar}) and the remaining terms, equal to zero.  The first 
equation (associated with the terms in the curly bracket) determines the 
amount of supersymmetry preserved by and the geometry of the calibrated 
surface.  The second equation is for the self-dual string soliton on the  
calibrated surface.  The amount of preserved supersymmetry can be determined 
by considering the supersymmetry projectors of the associated intersecting 
M5-brane configuration and of the added M2-brane.  One can also add an 
M2-brane to an intersecting M5-brane configuration without breaking 
additional supersymmetry, if the supersymmetry projectors associated with 
M5-branes yield the supersymmetry projector for the added M2-brane.  The 
scalar $\phi$ no longer satisfies the flat Laplace's equation, but satisfies 
an equation determined by the curved metric of the calibrated surface.

\section{Neutral String}

We consider the worldvolume soliton counterpart to the following target 
space configuration where M-wave travels along an longitudinal direction 
of an M5-brane:
\begin{equation}
\matrix{M5:&1&2&3&4&5\cr MW:& & & & &5}
\label{m5mw}
\end{equation}
When the $x^5$-direction is compactified, this configuration becomes 
a D0-brane in a D4-brane, where the D0-brane is interpreted as an instanton 
of the D4-brane worldvolume theory \cite{dou,wit}.  The worldvolume soliton 
counterpart to (\ref{m5mw}) has no active scalar and has non-vanishing 
${\cal H}_{0\tilde{m}\tilde{n}}$ and ${\cal H}_{5\tilde{m}\tilde{n}}$, 
where $\tilde{m},\tilde{n}=1,2,3,4$.  The nonzero component 
${\cal H}_{5\tilde{m}\tilde{n}}=F_{\tilde{m}\tilde{n}}$ is (anti-) self-dual 
as a two-form in four-dimensional Euclidean space and gives rise to a 
string-like soliton in the $x^5$-direction.  But such string-like soliton 
does not carry charge of the ${\cal H}$ field.  So, such soliton is called 
an instanton or a neutral string.  The corresponding bosonic worldvolume field 
Ansatz is
\begin{equation}
h_{0\tilde{a}\tilde{b}}=\pm h_{5\tilde{a}\tilde{b}}\equiv F_{\tilde{a}
\tilde{b}}, 
\label{neustrng}
\end{equation}
where $\tilde{a},\tilde{b}=1,...,4$ and all of the scalar fields 
$X^{a^{\prime}}$ are set to zero (hence, the induced metric is flat, $g_{mn}
=\eta_{mn}$).  To consider an instanton string solution on the noncommutative 
M5-brane worldvolume, one has to impose the boundary condition that $h_{mnp}$ 
is nonzero constant at infinity.  In order for the boundary condition to be 
compatible with the field Ansatz (\ref{neustrng}), one has to go to the 
infinite momentum frame (through infinite boost along the $x^5$-direction), 
in which the nonzero components of $h_{mnp}$ at infinity are given by 
Eq. (\ref{newbdcdtn}).  This boundary condition can be imposed on the field 
Ansatz (\ref{neustrng})  only for the negative sign choice in Eq. 
(\ref{neustrng}), i.e. only when $F_{\tilde{m}\tilde{n}}$ is anti-self-dual
\footnote{The infinite boost along the negative $x^{5}$-direction 
would lead to the boundary condition on $h_{abc}$ with opposite signs, 
and therefore select the self-dual $F_{\tilde{m}\tilde{n}}$ (i.e. the 
positive sign in Eq. (\ref{neustrng})).}.  
Then, the supersymmetry variation (\ref{susyvar}) of fermions takes the 
following form:
\begin{equation}
\hat{\delta}\Theta=-{1\over 2}\epsilon F^{\tilde{m}\tilde{n}}
(\gamma_{0\tilde{m}\tilde{n}}-\gamma_{5\tilde{m}\tilde{n}}).
\label{susyvarinstan}
\end{equation}
The resulting Killing spinor equation $\hat{\delta}\Theta=0$ can be satisfied, 
if we impose the following constraint on spinors:
\begin{equation}
\epsilon\gamma_0\gamma_5=\epsilon.
\label{spncnstinst}
\end{equation}

We show that a neutral string on a calibrated surface in the constant 
$h_{abc}$ field background also preserves supersymmetry.  For this purpose, 
it is convenient to introduce the light-cone coordinates:
\begin{equation}
u={1\over{\sqrt{2}}}(x^0-x^5),\ \ \ \ \ \ 
v={1\over{\sqrt{2}}}(x^0+x^5).  
\label{lccoord}
\end{equation} 
In terms of the light-cone coordinates, the boundary condition 
(\ref{newbdcdtn}) on the $h_{abc}$ field is
\begin{equation}
h_{u12}=-h_{u34}={\rm constant},
\label{lchfld}
\end{equation}
and the bosonic field Ansatz (\ref{neustrng}) with negative sign becomes
\begin{equation}
h_{u\tilde{a}\tilde{b}}:=\bar{F}_{\tilde{a}\tilde{b}}.
\label{newfldanz}
\end{equation}
From Eq. (\ref{relbetwnhH}), one can see that the only nonzero component of 
the corresponding ${\cal H}$ field is ${\cal H}_{u\tilde{m}\tilde{n}}=
\bar{F}_{\tilde{m}\tilde{n}}$.  The supersymmetry variation of fermions is
\begin{equation}
\tilde{\delta}\Theta=-{1\over 2}\epsilon\left[\bar{F}^{\tilde{m}\tilde{n}}
\gamma_{v\tilde{m}\tilde{n}}+\bar{F}^{\tilde{m}\tilde{n}}\partial_{\tilde{m}}
X^{I_1}\partial_{\tilde{n}}X^{I_2}\gamma_v\gamma_{I_1I_2}\right],
\label{susyvarcalneu}
\end{equation}
where $X^I$ are scalars associated with the calibrated surface and 
$\epsilon$ is the spinor for supersymmetry preserved by the 
calibrated surface, determined by setting the terms in the curly 
bracket of Eq. (\ref{calsfdsusyvar}) equal to zero.  
By imposing the spinor constraint $\epsilon\gamma^u=0$, one can set this 
supersymmetry variation to zero, making use of the relation 
$\gamma_v=\gamma^u$.  The possibility of having such supersymmetric 
configuration depends on whether the supersymmetry projectors of the 
intersecting M5-brane configuration associated with the calibrated surface 
yields the projector $\epsilon\gamma^u=0$ (or $\epsilon\gamma^0\gamma^5=
\epsilon$).  Had one used the boundary condition on $h_{abc}$ corresponding 
to the infinitely boosted frame along the negative $x^5$-direction, one 
would have had a self-dual instanton string on a calibrated surface with 
${\cal H}_{v\tilde{m}\tilde{n}}:=\bar{G}_{\tilde{m}\tilde{n}}$ and with the 
associated spinor constraint $\epsilon\gamma^v=0$.

\section{Traveling Wave on Calibrated Surface}

In this section we show that supersymmetric traveling wave on a 
calibrated surface exists even in the nonzero constant $h_{abc}$ 
field background.  Traveling wave on the M5-brane worldvolume is 
regarded as fluctuations in the shape of the calibrated surface.  
We consider the wave traveling in the $x^5$-direction.  For the 
purpose of studying traveling wave, it is convenient to work with  
the light-cone coordinates (\ref{lccoord}) and constant background 
$h_{abc}$ field of the form (\ref{lchfld}) in the infinite momentum frame.  
We expect that either purely left-moving or right-moving wave moving at 
the speed of light, i.e. the case when scalar fields $X^{a^{\prime}}$ 
do not depend on either $u$ or $v$, preserves supersymmetry.  We consider 
the case when $X^{a^{\prime}}$ are independent of $v$, i.e. $\partial_v
X^{a^{\prime}}=0$.  Then, the supersymmetry transformation (\ref{susyvar}) 
reduces to the following form:
\begin{eqnarray}
\hat{\delta}\Theta&=&-{1\over 2}\epsilon\left[{\rm det}(e^{-1})\partial_u
X^{a^{\prime}}\gamma^u\gamma_{a^{\prime}}\right.
\cr
& &\ \ \ -{1\over{3!}}{\rm det}(e^{-1})\partial_uX^{a^{\prime}_1}
\partial_{m_2}X^{a^{\prime}_2}\partial_{m_3}X^{a^{\prime}_3}
\gamma^{um_2m_3}\gamma_{a^{\prime}_1a^{\prime}_2a^{\prime}_3}
\cr
& &\ \ \ +{1\over{5!}}{\rm det}(e^{-1})\partial_uX^{a^{\prime}_1}
\partial_{m_2}X^{a^{\prime}_2}\cdots\partial_{m_5}X^{a^{\prime}_5}
\gamma^{um_2...m_5}\gamma_{a^{\prime}_1...a^{\prime}_5}
\cr
& &\ \ \ -h^{vm_2m_3}\partial_{m_2}X^{a^{\prime}_2}\partial_{m_3}
X^{a^{\prime}_3}\gamma_v\gamma_{a^{\prime}_2a^{\prime}_3}
\cr
& &\ \ \ \left.-{1\over 3}h^{vm_2m_3}\gamma_{vm_2m_3}\right].
\label{wavesusyvar}
\end{eqnarray}
This supersymmetry variation can be set to zero by imposing the spinor 
constraint $\epsilon\gamma^u=0$.  So, traveling wave in the constant 
$h_{abc}$ field background (\ref{lchfld}) with arbitrary dependence of 
$X^{a^{\prime}}$ on $u$ preserves supersymmetry.   
The analysis of the case with no dependence on $u$, i.e. $\partial_u
X^{a^{\prime}}=0$, is along the same line.  The corresponding 
supersymmetry projector is $\epsilon\gamma^v=0$ (or $\epsilon\gamma^0\gamma^5
=-\epsilon$).\\
\\
\noindent
{\large\bf Note Added} 

While this work was being completed, there appeared the paper \cite{mic} 
which has overlapping results with the section 3 of our paper but 
with slightly different derivation of the self-dual string soliton from 
ours.

\end{document}